\documentclass[aps,prd,reprint,superscriptaddress,10pt,a4paper,preprintnumbers,nofootinbib]{revtex4-2}
\usepackage{amsmath,amssymb,latexsym}
\usepackage{bm}
\usepackage{graphicx}
\usepackage{hepunits}
\usepackage{slashed}
\usepackage[svgnames]{xcolor}
\usepackage[normalem]{ulem}
\usepackage{booktabs}

\usepackage{textcomp}

\allowdisplaybreaks[4]

\begin{document}

\title{NNLO QCD calculation of hadron multiplicities in light-quark jets at lepton colliders}

\author{Bin Zhou}
\email{zb0429@sjtu.edu.cn}
\author{Jun Gao}
\email{jung49@sjtu.edu.cn}
\affiliation{
    State Key Laboratory of Dark Matter Physics, Shanghai Key Laboratory for Particle Physics and Cosmology, Key Laboratory for Particle Astrophysics and Cosmology (MOE), School of Physics and Astronomy, Shanghai Jiao Tong University, Shanghai 200240, China
 }

\begin{abstract}
We present the calculation of next-to-next-to-leading-order (NNLO) QCD corrections to hadron multiplicities in light-quark jets at lepton colliders, employing the ``projection-to-Born" (P2B) method implemented in the FMNLO program. Taking the next-to-leading-order result as an example, we rigorously establish the validity of our P2B-based calculation. We then present NNLO predictions for the normalized asymmetry $D_{K^{-}}$ between hadron and antihadron production in light-quark jets and compare them with SLD data. 
We find that a suppression of these SLD measurements relative to NPC23 predictions for $D_{K^{-}}$ emerges in the intermediate $z_h$ domain ($0.2 \lesssim z_h \lesssim 0.7$).
We expect that incorporating these SLD data into global QCD fits will enable improved determination of fragmentation functions.
\end{abstract}

\maketitle

\section{Introduction}
In QCD, the production of identified hadrons in hard-scattering processes involves a two-stage mechanism: first, the perturbative production of partons (quarks or gluons), followed by the nonperturbative transition of these partons into hadrons. This hadronization process is characterized by fragmentation functions (FFs), which quantify the probability for a parton to produce a hadron carrying a fraction of its momentum~\cite{Berman:1971xz,Field:1976ve,FIELD19781}.
Besides their fundamental role in understanding QCD color confinement, FFs are crucial for probing the nucleon substructure -- heightened in importance by the advent of the precision era in high-energy nuclear physics enabled by electron-ion colliders~\cite{Accardi:2012qut,Anderle:2021wcy}.

Given that FFs encode nonperturbative QCD dynamics, they cannot currently be calculated from first principles of QCD and are instead extracted via global fits to experimental data from $e^+e^-$, lepton-hadron, and hadron-hadron collisions. Perturbative QCD corrections to single-inclusive annihilation (SIA) at lepton colliders have been calculated up to next-to-next-to-next-to-leading-order ($\text{N}^3\text{LO}$) in Refs.~\cite{Rijken:1996vr,Rijken:1996npa,Rijken:1996ns,Mitov:2006wy,Blumlein:2006rr,he2025singleinclusivehadronproductionelectronpositron}.  For semi-inclusive deep inelastic scattering (SIDIS) in lepton-hadron collisions, the corresponding cross sections at next-to-next-to-leading-order (NNLO) have been calculated in Refs.~\cite{Altarelli:1979kv,Nason:1993xx,Furmanski:1981cw,Graudenz:1994dq,deFlorian:1997zj,deFlorian:2012wk,Goyal:2023zdi,Bonino:2024wgg,Bonino:2024qbh, Goyal:2024tmo,Goyal:2024emo,Bonino:2025tnf}. Approximate NNLO and $\text{N}^3\text{LO}$ corrections have also been derived from the expansion of the threshold resummation expressions~\cite{Abele:2021nyo, Abele:2022wuy}.
Equipped with perturbative QCD calculations and a wealth of experimental measurements, comprehensive global fits incorporating diverse data samples have become feasible for extracting FFs. The representative efforts at next-to-leading-order (NLO) accuracy can be found in the works of 
DSS~\cite{deFlorian:2007aj,deFlorian:2007ekg,deFlorian:2014xna,deFlorian:2017lwf}, HKNS~\cite{Borsa:2021ran}, AKK~\cite{Albino:2008fy}, NNFF~\cite{Bertone:2018ecm}, MAPFF~\cite{Khalek:2021gxf}, JAM~\cite{Moffat:2021dji}, and NPC23~\cite{Gao:2024nkz,Gao:2024dbv,gao2025globalanalysisfragmentationfunctions}.
Furthermore, determinations of FFs have been made at NNLO with SIA data only~\cite{Bertone:2017tyb,Soleymaninia:2018uiv},
at approximate NNLO with SIA and SIDIS data~\cite{Borsa:2022vvp,AbdulKhalek:2022laj}, and most recently at full NNLO accuracy in QCD for charged pions
and kaons, using a global analysis of SIA and SIDIS data~\cite{gao2025fragmentationfunctionschargedhadrons}.

In $e^+e^-$ collisions, numerous SIA measurements have been performed by various experiments at different center-of-mass energies~\cite{TASSO:1988jma,TPCTwoGamma:1988yjh,OPAL:1994zan,ALEPH:1994cbg,DELPHI:1998cgx,SLD:2003ogn,OPAL:2002isf,DELPHI:2000ahn}. 
For instance, analyses based on SLD measurements at the $Z$-pole have been presented for inclusive charged particles as well as identified $\pi^\pm$, $K^\pm$, and $p/\bar{p}$ production in hadronic $Z^0$ decays. These studies distinguish production originating from light ($u, d, s$), charm ($c$), and bottom ($b$) primary quark flavors.
Beyond measuring hadron multiplicity distributions in SIA, experiments also provide differential production rates for hadrons and antihadrons specifically in light-quark jets~\cite{SLD:2003ogn} (detailed in subsequent sections). Similar to analyses of hadron energy fraction distributions in the forward region of SIA processes~\cite{Zhou:2024cyk}, studying hadron multiplicities in light-quark jets offers a valuable method for the separation of light-quark and antiquark FFs.
Consequently, incorporating this dataset into global FF fits is expected to enhance the precision of the extracted FFs, necessitating accurate theoretical calculations of the corresponding observable.

In this work, we focus on the NNLO calculations of hadron multiplicity distributions in light-quark jets. Leveraging the capability of the FMNLO program~\cite{Liu:2023fsq,Gao:2024dbv,Zhou:2024cyk} to calculate hadron multiplicities for both NNLO-accurate SIA processes and NLO-accurate three-jet production processes~\cite{Zhou:2024cyk}, we employ the ``projection-to-Born" (P2B) method~\cite{Cacciari:2015jma, Dreyer:2016oyx,Currie:2018fgr,Dreyer:2018qbw,Gehrmann:2018odt,Chen:2021isd} to derive NNLO corrections for hadron multiplicities in light-quark jets. 
FMNLO is a program for automated and fast calculations of fragmentation cross sections at NLO for arbitrary 
processes via MG5\_aMC@NLO~\cite{Alwall:2014hca,Frederix:2018nkq}. It is based on a hybrid scheme combining the phase-space slicing and local subtraction methods.
The updated version, FMNLOv2.1, extends this capability to compute NNLO corrections for hadron multiplicities in SIDIS, SIA, and Higgs boson decays to gluons~\cite{Zhou:2024cyk}.
For these NNLO calculations within FMNLO, the convolution of analytic coefficient functions ~\cite{Rijken:1996vr,Rijken:1996npa,Rijken:1996ns,Mitov:2006wy,Blumlein:2006rr} with FFs is performed via numerical Monte Carlo integration.
Building on these results for hadron multiplicities in light-quark jets, we further analyze differential production rate asymmetries (specifically, the normalized difference $D_h$ defined in Eq.~\eqref{eq:Dh}) between hadrons and antihadrons in light-quark jets.

This paper is organized as follows. Section~\ref{sec:methods} introduces the theoretical formalism, including the definition of hadron multiplicities in light-quark jets and their computation using the P2B method. In Sec.~\ref{sec:results}, we first demonstrate the validity of our P2B-based calculation and then present NNLO numerical results for the asymmetries between hadron and antihadron production in light-quark jets, comparing them with SLD data. Conclusions and prospects are summarized in Sec.~\ref{sec:conclusion}. 
The Appendix provides a brief introduction to our NNLO calculations for the hadron energy fraction distribution in light-quark jets, as implemented in FMNLO and used in this study.

\section{Methods}
\label{sec:methods}
In this section, we present the calculation of hadron multiplicities in light-quark jets, which can be used to distinguish between quark and antiquark flavor fragmentation functions. We start by reviewing the factorization formula for the SIA process. The spin-averaged SIA differential cross section in the energy fraction $z_h\equiv 2E_h/Q$ carried by the identified hadron at a given center-of-mass energy of $\sqrt{s} = Q$ is written as
\begin{equation}
 \frac{d\sigma^h}{d z_h}= \sum_{i=q,\bar{q},g} \frac{d\sigma_i}{dy} \otimes D_i^h(z_h)
 \label{eq:11}
\end{equation}
Here, $D_i^h(x)$ denotes the fragmentation function of parton $i$ into hadron $h$. The symbol $\otimes$ denotes the standard convolution integral, defined as 
\begin{equation}\label{convolution}
f \otimes g(z)
=\int^1_0 dx_1 \int^1_0 dx_2 f(x_1) g(x_2) \delta(z - x_1x_2) \,,
\end{equation}
In Eq.~\eqref{eq:11}, $\frac{d\sigma_i}{dy}$ is the differential cross section in the energy fraction $y$ carried by the parton $i$ (quark $q$, antiquark $\bar{q}$, and gluon $g$). It can be formally expressed as
\begin{equation}
 \frac{d\sigma_i}{d y}= \int d\text{PS} \, \frac{d\sigma_i}{d\text{PS}} \,\delta(y-\frac{2E_i}{\sqrt s})
 \label{eq:1}
\end{equation}
where $E_i$ is the energy of parton $i$ and $d\text{PS}$ denotes the final-state phase-space measure. The analytic results for $\frac{d\sigma_i}{dy}$ can be expressed in terms of the leading-order (LO) inclusive cross section $\sigma_i^{(0)}$ for parton $i$ production and the SIA perturbative coefficient functions. These results have been calculated up to $\text{N}^3\text{LO}$~\cite{Rijken:1996vr,Rijken:1996npa,Rijken:1996ns,Mitov:2006wy,Blumlein:2006rr,he2025singleinclusivehadronproductionelectronpositron}. 
In $\sigma_i^{(0)}$, the vector- and axial-vector coupling constants of the $Z$-boson to the lepton can be replaced by the corresponding left- and right-handed couplings. This replacement enables one to readily obtain the hadron energy fraction distribution for a polarized initial-state electron by scaling the left-handed couplings by $\sqrt{1 - P}$ and the right-handed couplings by $\sqrt{1 + P}$, where $P = -1$ corresponds to a left-handed electron and $P = 1$ to a right-handed electron.

We now address the calculation of hadron multiplicities in light-quark jets, corresponding to the quark-tagged hemisphere in experiments~\cite{SLD:2003ogn}.
The hadron energy fraction distribution in the quark-tagged hemisphere can be expressed similarly to Eq.~\eqref{eq:11}, but with $\frac{d\sigma_i}{dy}$ modified as
\begin{equation}
 \frac{d\sigma_i}{d y}= \int d\text{PS}  \frac{d\sigma_i}{d\text{PS}} \delta(y-\frac{2E_i}{\sqrt s}) \Theta(\text{cos}\theta_{\vec{h}\vec{n}}) F(\text{cos}\theta_{\vec{n}})
 \label{eq:1}
\end{equation}
where $\vec{n}$ denotes the thrust axis (defined with 
$n_z=\text{cos}\theta_{\vec{n}}>0$ aligned with the electron beam), and $\theta_{\vec{h}\vec{n}}$ is the angle of hadron $h$ relative to the thrust axis in the center-of-mass frame. Note that the hadron’s direction is aligned with the momentum of its parent parton $i$.
$F(\text{cos}\theta_{\vec{n}})$ denotes an optional selection criterion applied to the thrust axis orientation. 
For left-handed (right-handed) electron beam polarization, the quark-tagged hemisphere contains tracks with positive (negative) momentum projection along the signed thrust axis. The remaining tracks in each event are defined to be in the antiquark-tagged hemisphere. Focusing on left-handed polarization, we require final-state hadrons to satisfy $\text{cos}\theta_{\vec{h}\vec{n}}>0$.

Using the FMNLO program~\cite{Liu:2023fsq} interfaced to MG5\_aMC@NLO~\cite{Alwall:2014hca,Frederix:2018nkq}, we can compute the hadron energy fraction distribution in the quark-tagged hemisphere at NLO. FMNLO employs a hybrid phase-space slicing and local subtraction scheme to enable automated NLO calculations of fragmentation cross sections, including SIA. By implementing the angular constraints $\Theta(\cos\theta_{\vec{h}\vec{n}})$ and $F(\cos\theta_{\vec{n}})$ within the SIA module (as detailed in FMNLO and Appendix~\ref{Appendix:benchmark}), we thus derive the hadron energy fraction distribution in the quark-tagged hemisphere.
This result serves as the benchmark for validating the P2B method discussed below.

In the NNLO SIA calculations using FMNLO, the convolution of analytic coefficient functions with FFs is performed via numerical Monte Carlo integration. The implementation based on the analytic coefficient functions~\cite{Rijken:1996vr,Rijken:1996npa,Rijken:1996ns,Mitov:2006wy,Blumlein:2006rr} cannot inherently determine the thrust axis or enforce angular cuts. Therefore, direct modification of the FMNLO NNLO SIA implementation to include angular constraints is infeasible. 
To address this limitation, we employ the P2B method~\cite{Cacciari:2015jma, Dreyer:2016oyx,Currie:2018fgr,Dreyer:2018qbw,Gehrmann:2018odt,Chen:2021isd}.
Decomposing Eq.~\eqref{eq:1} with a cutoff $\tau$, we write
\begin{align}
\label{eq:2}
     &\frac{d\sigma_i}{d y}= \int d\text{PS}  \frac{d\sigma_i}{d\text{PS}} \delta(y-\frac{2E_i}{\sqrt s}) \Theta(\text{cos}\theta_{\vec{h}\vec{n}}) \\ \notag
    &\times\big[ \Theta(\tau-\tau_0)+\Theta(\tau_0-\tau)\big] F(\text{cos}\theta_{\vec{n}})\\\notag
     =& \int d\text{PS}  \frac{d\sigma_i}{d\text{PS}} \delta(y-\frac{2E_i}{\sqrt s}) \Theta(\text{cos}\theta_{\vec{h}\vec{n}})\Theta(\tau-\tau_0)F(\text{cos}\theta_{\vec{n}}) \\\notag
     +& \int d\text{PS}  \frac{d\sigma_i}{d\text{PS}} \delta(y-\frac{2E_i}{\sqrt s}) \Theta(\text{cos}\theta_{\vec{h}\vec{n}})\Theta(\tau_0-\tau)F(\text{cos}\theta_{\vec{n}})
\end{align}
where we can choose $\tau\equiv1-T$, with $T$ being the thrust variable.
For $\tau_0\ll1$, the final state approximates two back-to-back jets, with $\text{cos}\theta_{\vec{h}\vec{n}}\approx\text{sign}[\text{cos}\theta_{\vec{h}}]$ and $\text{cos}\theta_{\vec{n}}\approx\lvert \text{cos}\theta_{\vec{h}} \rvert$. The soft-collinear integration then becomes
\begin{equation}
    \begin{split}
&\int d\text{PS}  \frac{d\sigma_i}{d\text{PS}} \delta(y-\frac{2E_i}{\sqrt s}) \Theta(\text{cos}\theta_{\vec{h}\vec{n}})\Theta(\tau_0-\tau)F(\text{cos}\theta_{\vec{n}})\quad\quad\quad\quad\quad\\
&=\int d\text{PS}  \frac{d\sigma_i}{d\text{PS}} \delta(y-\frac{2E_i}{\sqrt s}) \Theta(\text{cos}\theta_{\vec{h}})\Theta(\tau_0-\tau)F(\text{cos}\theta_{\vec{h}})+ \mathcal{O}(\tau_0)\\
&=\int d\text{PS}  \frac{d\sigma_i}{d\text{PS}} \delta(y-\frac{2E_i}{\sqrt s}) \Theta(\text{cos}\theta_{\vec{h}})F(\text{cos}\theta_{\vec{h}})-\int d\text{PS}\\
&  \times\frac{d\sigma_i}{d\text{PS}} \delta(y-\frac{2E_i}{\sqrt s}) \Theta(\text{cos}\theta_{\vec{h}})\Theta(\tau-\tau_0)F(\text{cos}\theta_{\vec{h}}) + \mathcal{O}(\tau_0)
\label{eq:3}
    \end{split}
\end{equation}
where we employ the substitution $1-\Theta(\tau-\tau_0)$ for $\Theta(\tau_0-\tau)$. 
Substituting Eq.~\eqref{eq:2} and Eq.~\eqref{eq:3} into Eq.~\eqref{eq:11}, we have 
\begin{equation}
    \begin{split}
     &\frac{d\sigma^h}{d z_h} =  \sum_{i=q,\bar{q},g}\int d\text{PS}  \frac{d\sigma_i}{d\text{PS}} \delta(y-\frac{2E_i}{\sqrt s}) \Theta(\text{cos}\theta_{\vec{h}})F(\text{cos}\theta_{\vec{h}}) \\
     &\otimes D_i^h(z_h) + \sum_{i=q,\bar{q},g}\int d\text{PS}  \frac{d\sigma_i}{d\text{PS}} \delta(y-\frac{2E_i}{\sqrt s}) \Theta(\tau-\tau_0)\\
     &\times\big[ \Theta(\text{cos}\theta_{\vec{h}\vec{n}}) F(\text{cos}\theta_{\vec{n}})-\Theta(\text{cos}\theta_{\vec{h}})F(\text{cos}\theta_{\vec{h}}) \big]\\
     &\otimes D_i^h(z_h)+ \mathcal{O}(\tau_0).
     \label{eq:44}
    \end{split}
\end{equation}
valid in the limit $\tau_0\to 0$. In this limit, the expression simplifies to:
\begin{equation}
    \begin{split}
     &\frac{d\sigma^h}{d z_h} =  \sum_{i=q,\bar{q},g}\int d\text{PS}  \frac{d\sigma_i}{d\text{PS}} \delta(y-\frac{2E_i}{\sqrt s}) \Theta(\text{cos}\theta_{\vec{h}})F(\text{cos}\theta_{\vec{h}}) \\
     &\otimes D_i^h(z_h) + \sum_{i=q,\bar{q},g}\int d\text{PS}  \frac{d\sigma_i}{d\text{PS}} \delta(y-\frac{2E_i}{\sqrt s}) \\
     &\times\big[ \Theta(\text{cos}\theta_{\vec{h}\vec{n}}) F(\text{cos}\theta_{\vec{n}})-\Theta(\text{cos}\theta_{\vec{h}})F(\text{cos}\theta_{\vec{h}}) \big]\otimes D_i^h(z_h).
     \label{eq:4}
    \end{split}
\end{equation}
which is the expression derived via the P2B method. Implementation of the first term in Eq.~\eqref{eq:4} at NNLO accuracy via analytic coefficient functions is straightforward (see Appendix~\ref{Appendix:first}). The second term, however, receives contributions exclusively from final states with three or more jets. Crucially, the hadron energy fraction distributions for these three-jet processes -- subject to the kinematic cuts $\Theta(\text{cos}\theta_{\vec{h}\vec{n}}) F(\text{cos}\theta_{\vec{n}})$ or $\Theta(\text{cos}\theta_{\vec{h}})F(\text{cos}\theta_{\vec{h}})$ -- are infrared-safe observables. Consequently, this term can be calculated using the FMNLO program, as described in Appendix~\ref{Appendix:second}. 
Finally, it should be noted that the program based on the P2B method can compute not only NNLO corrections but also NLO corrections to the hadron energy fraction distribution in the quark-tagged hemisphere.

\section{Numerical Results}
\label{sec:results}
In this section, we present numerical predictions for hadron multiplicities in light-quark jets. The SLD experiment at SLAC conducted measurements for the SIA process. In their SIA study~\cite{SLD:2003ogn}, momentum spectra for light charged hadrons are measured using a polarized electron beam colliding with a positron beam. Unless otherwise specified, the lepton collider center-of-mass energy is set to $\sqrt{s}=91.18 \,\text{GeV}$ and 
the magnitude of the $e^-$ beam polarization, defined as
\begin{align}
P_e = \frac{N^{-} - N^{+}}{N^{-} + N^{+}},
\end{align}
is fixed at $0.73$. Here, $N^{-}$ denotes the total number of events with left-handed electrons, and $N^{+}$ denotes the total number of events with right-handed electrons. 
At the parton level, only the three light-quark flavors ($u$, $d$, and $s$) are considered for the electroweak vertices. Events are accepted if $\text{cos}\theta_{\vec{h}\vec{n}}>0$ and $0.15<n_z<0.71$. Therefore, the measured hadron multiplicities can be formally expressed as  
\begin{align}
\label{eq:10}
R_{h}^{q}&\equiv\frac{1}{\sigma} \frac{d\sigma^h}{d z_h} = \frac{1}{\sigma}\sum_{i=q,\bar{q},g}\int d\text{PS}  \frac{d\sigma_i}{d\text{PS}} \delta(y-\frac{2E_i}{\sqrt s}) \\\notag
 &\times\Theta(\text{cos}\theta_{\vec{h}\vec{n}}) \Theta(n_z-0.15)\,\Theta(0.71-n_z) \otimes D_i^h(z_h)
\end{align}
where $q$ in the symbol $R_{h}^{q}$ represents light-flavor quark jets (quark-tagged hemisphere) and $h$ represents light charged hadrons. For brevity, we focus on the $h=K^-$ spectra. $\sigma$ is the inclusive total cross section for SIA incorporating the specified additional kinematic cuts $F(\text{cos}\theta_{\vec{n}})=\Theta(n_z-0.15)\,\Theta(0.71-n_z)$, which could be calculated using the P2B method. Similar to Eq.~\eqref{eq:4}, the expression for $\sigma$ based on the P2B method can be written as
\begin{equation}
    \begin{split}
     \sigma &= \int d\text{PS}  \frac{d\sigma}{d\text{PS}} F(\text{cos}\theta_{q}) \\ & + \int d\text{PS}  \frac{d\sigma}{d\text{PS}} \big[ F(\text{cos}\theta_{\vec{n}})-F(\text{cos}\theta_{q}) \big],
     \label{eq:44}
    \end{split}
\end{equation}
where $\theta_{q}$ is the angle between the outgoing quark $q$ and the incoming electron. $\sigma$ can also be calculated separately using other programs~\cite{NNLOJET:2025rno}, showing good agreement with the results obtained from the P2B method.
Before comparing with the data from SLD, we first demonstrate the validity of our P2B-based calculation.

\begin{figure}[t!]
\centering
\includegraphics[width=0.4\textwidth]{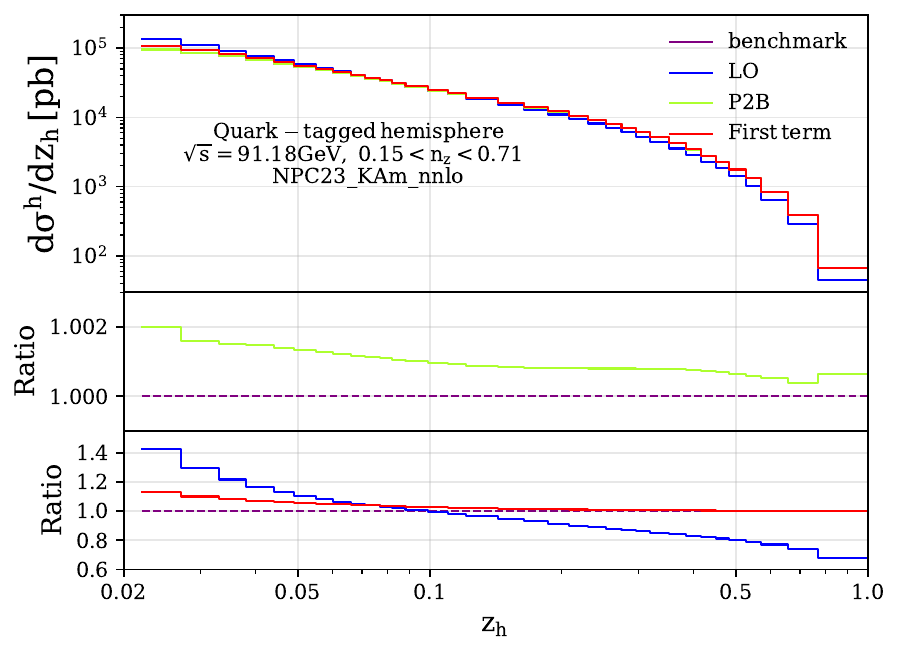}
\vspace{-3ex}
\caption{Comparison of the NLO results obtained using the P2B method with the benchmark calculation of the hadron energy fraction distribution in the quark-tagged hemisphere. The middle panel displays the ratios of the P2B predictions to the benchmark, while the bottom panel shows the ratio of the LO prediction and of the first term in Eq.~\eqref{eq:4} to the benchmark.}
\label{fig:benchmark}
\end{figure}
Figure~\ref{fig:benchmark} contains three panels comparing the NLO results obtained using the P2B method with the benchmark results for the hadron energy fraction distribution in the quark-tagged hemisphere. The middle panel displays the ratios of the P2B predictions to the benchmark. For these calculations, we employ the NNLO FF sets from Ref.~\cite{gao2025fragmentationfunctionschargedhadrons}.
As shown in Fig.~\ref{fig:benchmark}, excellent agreement between the benchmark and the P2B results (calculated using Eq.~\eqref{eq:4}) is observed across the entire kinematic range, as evidenced by the middle panel. Specifically, deviations in all bins are below $0.2\%$, primarily attributed to Monte Carlo uncertainties, which validates the P2B method implementation at NLO.
The bottom panel shows the ratio of the LO prediction and the first term in Eq.~\eqref{eq:4} to the benchmark. We observe that the NLO correction first decreases and then increases with rising $z_h$, reaching up to $42\%$. The ratio of the first term alone decreases monotonically with increasing $z_h$, indicating that the contribution of the second term to the P2B result diminishes as $z_h$ grows, reaching a maximum reduction of $12\%$.

\begin{figure}[h!]
\centering
\includegraphics[width=0.45\textwidth]{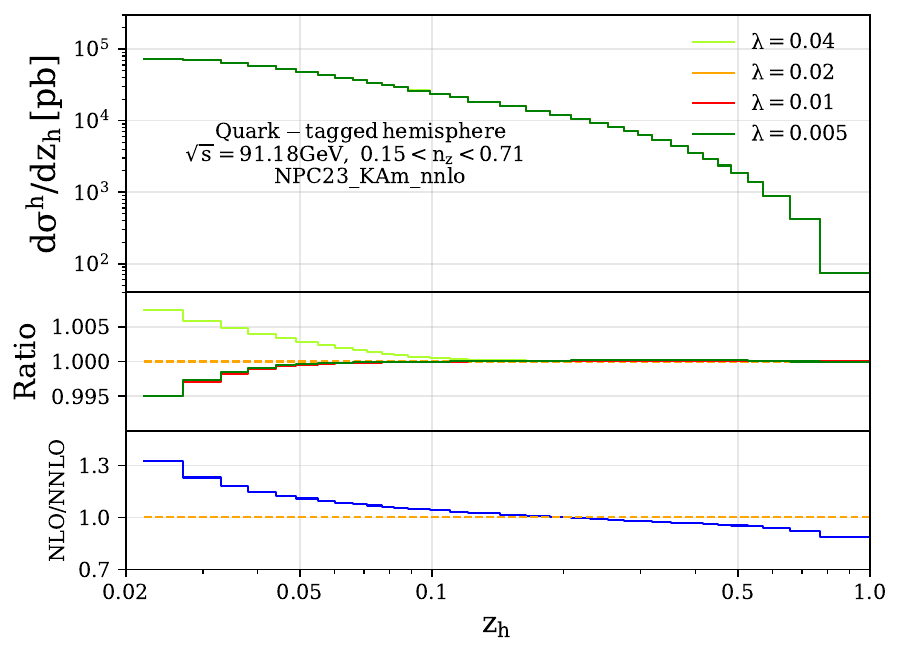}
\vspace{-3ex}
\caption{Comparison of the NNLO predictions for the hadron energy fraction distribution in the quark-tagged hemisphere for $\lambda$ values ranging from $0.005$ to $0.04$. The middle and lower panels show, respectively, the ratios of predictions at $\lambda=0.005$, $0.01$, and $0.04$ to the $\lambda=0.02$ reference, and the ratios of NLO results to NNLO ones at $\lambda=0.02$.
}
\label{fig:cut}
\end{figure}
We now turn to demonstrate the independence of our predictions for the hadron energy fraction distribution in the quark-tagged hemisphere from the choice of the cutoff parameter $\lambda$. Beyond LO, FMNLO introduces a cutoff parameter $\lambda$ to partition the phase space into unresolved and resolved collinear regions, as detailed in Ref.~\cite{Liu:2023fsq}. Since the P2B result at NLO incorporates only the LO contribution from three-jet production, which is independent of $\lambda$, we demonstrate in Fig.~\ref{fig:cut} the independence of our NNLO predictions from the choice of $\lambda$.
The upper panel shows the NNLO predictions of the differential cross sections as a function of $z_h$ for several choices of $\lambda$ ranging from $0.005$ to $0.04$. The middle and lower panels display, respectively, the ratios of the predictions obtained with $\lambda=0.005$, $0.01$, and $0.04$ to those with $\lambda=0.02$, and the ratios of the NLO results to the NNLO result with $\lambda=0.02$.
Owing to the large cancellation between 
$\Theta(\cos\theta_{\vec{h}\vec{n}}) F(\cos\theta_{\vec{n}}) - \Theta(\cos\theta_{\vec{h}})F(\cos\theta_{\vec{h}})$, our predictions for three-jet production exhibit a residual dependence on the cutoff parameter. However, the resulting variations across different $\lambda$ values are confined to within $1\%$, which has a negligible impact on the final physical observables compared to the $\sim 30\%$ NNLO correction. As evident in the middle panel, significant variations for different $\lambda$ values occur only at small $z_h$, a region typically excluded from fitting procedures.

Compared to hadron multiplicities in light-quark jets, we are more interested in the physical observable $D_h$, formally defined as the normalized asymmetry between hadron and antihadron production in light-quark jets:
\begin{align}
D_h=\frac{R_{h}^{q}-R_{\bar{h}}^{q}}{R_{h}^{q}+R_{\bar{h}}^{q}}
\label{eq:Dh}
\end{align}
where $\bar{h}$ denotes the antihadron corresponding to $h$.  
To address whether a primary quark-initiated jet (e.g., $u$ quark) produces more hadrons with valence quarks (e.g., $\pi^+$, $K^+$, $p$) than those without (e.g., $\pi^-$, $K^-$, $\bar{p}$), the SLD analysis~\cite{SLD:2003ogn} excluded antiquark contributions in the quark-tagged hemisphere, referred to as the ``Pure" result for clarity. 
The relationship between the ``Pure" result (excluding antiquarks) and the ``Mixing" result  (including antiquarks) is governed by the effective quark purity, which depends on the polar angle and was estimated from simulations to have an average value of $0.72$ for the selected track sample~\cite{SLD:2003ogn}.
It is crucial to emphasize that the current FMNLO implementation directly calculates only the ``Mixing" results, while the ``Pure" results are obtained from the calculated ``Mixing" ones using a matrix transformation based on the effective quark purity. 

\begin{figure}[h!]
\centering
\includegraphics[width=0.45\textwidth]{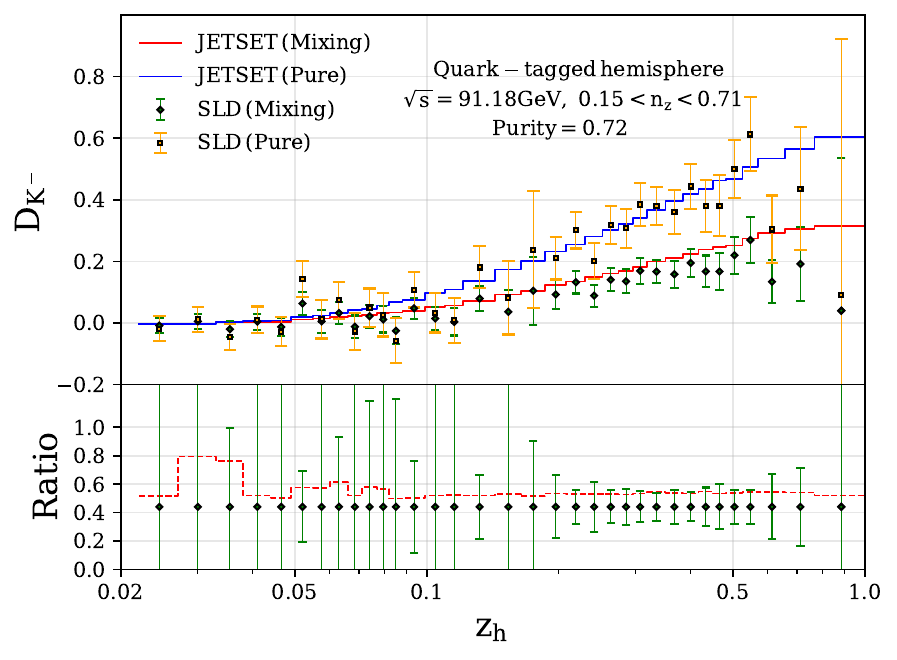}
\vspace{-3ex}
\caption{
Comparison of the normalized difference ($D_{K^{-}}$) between hadron and antihadron production in light-quark jets from SLD measurements and JETSET predictions. ``Pure" denotes results excluding antiquark contributions in the quark-tagged hemisphere, while ``Mixing" includes these contributions. The error bars include statistical and systematic uncertainties. The lower panel shows the ratio to the corresponding ``Pure" predictions.}
\label{fig:0310017_D_K_Purity}
\end{figure}

To validate this transformation between the ``Pure" and ``Mixing" results, Fig.~\ref{fig:0310017_D_K_Purity} compares the SLD measurements with JETSET predictions for both the ``Pure" and ``Mixing" cases of $D_{K^{-}}$. The error bars represent the combined statistical and systematic uncertainties, and the lower panel shows the ratio to the corresponding ``Pure" predictions.
It is important to note that the SLD ``Mixing" data for $D_{K^{-}}$ were obtained by applying a matrix transformation based on the effective quark purity to the original ``Pure" data, while the JETSET prediction was generated directly using the JETSET model~\cite{Sjostrand:1993yb}.
For the ``Pure" $D_{K^{-}}$ case, Fig.~\ref{fig:0310017_D_K_Purity} shows that the predictions of the JETSET model are consistent with the SLD data, as discussed in detail in Ref.~\cite{SLD:2003ogn}. In the ``Mixing" case, the JETSET predictions generally fall within the SLD experimental uncertainties. However, deviations beyond the error bars are observed in a few data bins, notably in the range $0.3 \lesssim z_h \lesssim 0.5$.
This discrepancy arises because the effective quark purity predicted by JETSET differs from the experimental value of $0.72$. As seen in the lower panel, the effective quark purity from JETSET is higher than $0.72$, but remains within the error bars. Therefore, in subsequent analyses, we use the ``Mixing" data from SLD as the baseline for studying the mixed scenario.

\begin{figure}[h!]
   \centering
   \includegraphics[width=0.45\textwidth]{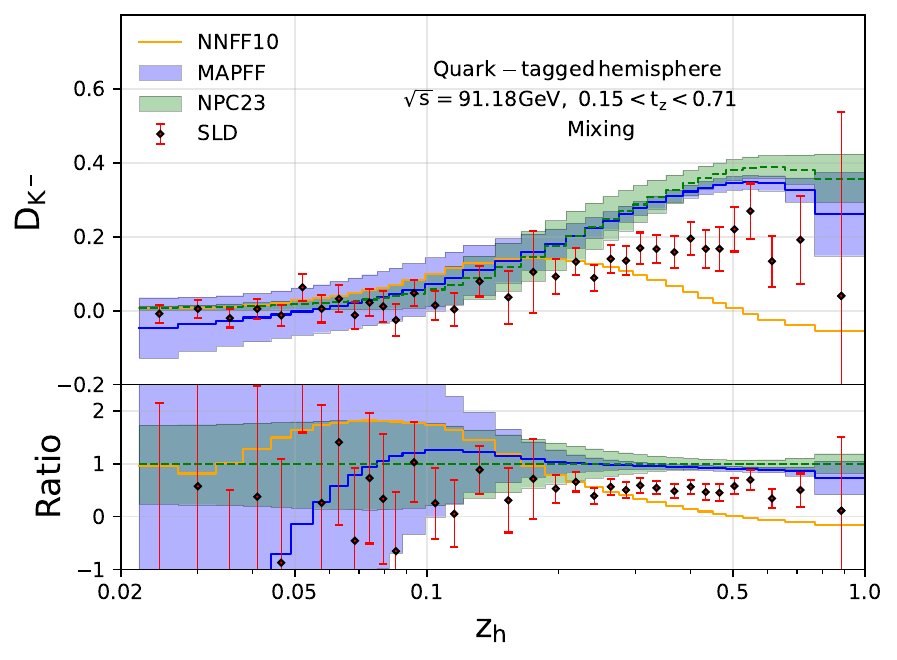}
   \vspace{-3ex}
   \caption{\label{fig:0310017_nnlo_D_Hessian}Comparison of $D_{k^-}$ predictions at NNLO between the results obtained from different fragmentation function sets and the data from SLD at the $Z$-pole. The lower panel displays the ratio to the NPC23 predictions. The error bands indicate their one-$\sigma$ uncertainties, and the error bars include statistical and systematic uncertainties.}
\end{figure}

Figure~\ref{fig:0310017_nnlo_D_Hessian} presents our NNLO predictions for the ``Mixing" $D_{K^{-}}$ observable, comparing results from various fragmentation function sets~\cite{gao2025fragmentationfunctionschargedhadrons,Bertone:2017tyb,AbdulKhalek:2022laj} with the data from SLD at the $Z$-pole. 
The lower panel shows the ratio to the NPC23 predictions, with error bands representing their one-$\sigma$ uncertainties for the FF sets~\cite{gao2025fragmentationfunctionschargedhadrons}, 
while error bars represent the combined statistical and systematic uncertainties of the data.
Owing to the lack of constraints from SIDIS data, the uncertainties for the NNFF10 results are too large and are, therefore, not shown in Fig.~\ref{fig:0310017_nnlo_D_Hessian}. All FFs used here are at NNLO accuracy.

We first analyze the impact of different FF sets on the NNLO results. Figure~\ref{fig:0310017_nnlo_D_Hessian} shows that the MAPFF results exhibit good agreement with the NPC23 predictions, considering their respective uncertainties. Meanwhile, the NNFF10 results align with the NPC23 results for $z_h<0.2$ but show progressively larger deviations as $z_h$ exceeds $0.2$. 
When comparing theoretical predictions with experimental data, the SLD results fall within the NPC23 uncertainty bands at low and high $z_h$, while systematically lying below theoretical values in the intermediate region ($0.2 \lesssim x \lesssim 0.7$).
This suppression may originate from either an overestimated fragmentation function for strange quarks producing $K^-$ or an underestimated fragmentation function for up quarks producing $K^+$. To enhance theoretical precision, incorporating this dataset into global fits is essential for extracting more accurate fragmentation functions.
\begin{figure}[h!]
   \centering
   \includegraphics[width=0.45\textwidth]{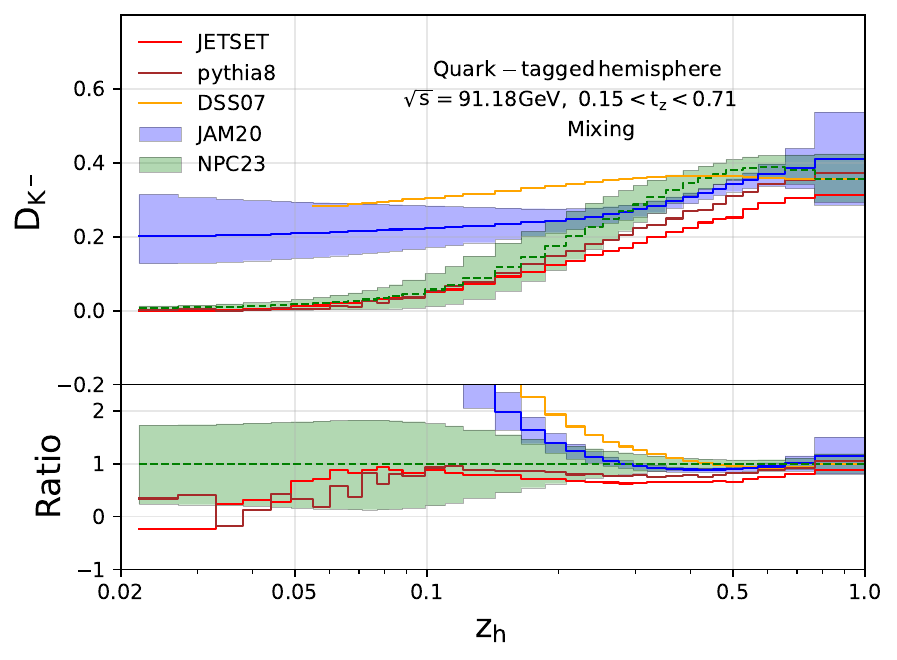}
   \vspace{-3ex}
   \caption{\label{fig:0310017_nnlo_D_Hessian_NLO}Comparison of our NNLO predictions from various FFs at NLO, JETSET, and Pythia8 models to the NPC23 results from the FFs at NNLO. The lower panel displays the ratio to the NPC23 results. The error bands indicate their one-$\sigma$ uncertainties.}
\end{figure} 

Figure~\ref{fig:0310017_nnlo_D_Hessian_NLO} compares our NNLO predictions from various FFs at NLO~\cite{Moffat:2021dji,deFlorian:2007aj,deFlorian:2007ekg,deFlorian:2014xna,deFlorian:2017lwf}, JETSET~\cite{Sjostrand:1993yb}, and Pythia8 models~\cite{Bierlich:2022pfr} against NPC23 results from the FFs at NNLO. 
Since no uncertainties for the DSS FFs are provided publicly, the error band for the DSS prediction is not available.
It can be observed in Fig.~\ref{fig:0310017_nnlo_D_Hessian_NLO} that the JAM and DSS results demonstrate good agreement with NPC23 in the high-$z_h$ region but exhibit significant discrepancies at low $z_h$. Conversely, JETSET and Pythia8 predictions remain within NPC23 uncertainties at low $z_h$. As $z_h$ increases through the intermediate range ($0.3 \lesssim z_h \lesssim 0.6$), however, the results from both models fall outside the NPC23 bands, with Pythia8 results showing smaller deviations than JETSET ones. Finally, at high $z_h$, JETSET and Pythia8 predictions agree with NPC23 within uncertainties.
\section{Conclusion}
\label{sec:conclusion}

In this work, we have presented the NNLO QCD calculation of light charged hadron multiplicities in light-quark jets at lepton colliders. This achievement was realized using the FMNLO framework combined with the P2B method.

The validity of our computational approach was rigorously established at NLO through excellent agreement between the NLO results obtained using the P2B method and the benchmark calculation of the hadron energy fraction distribution in the quark-tagged hemisphere. This successful validation provides a solid foundation for extending the P2B framework to NNLO accuracy.

Applying this validated NNLO formalism, 
we computed the normalized asymmetry $D_{K^{-}}$ for the ``Mixing" scenario in light-quark jets. Comparisons of our predictions, derived using various FF sets, with SLD data yielded significant findings.
We observe that the MAPFF results are in good agreement with the NPC23 ones across the entire $z_h$ range, considering their respective uncertainties, while the NNFF10 results 
align with the NPC23 ones only for $z_h<0.2$.
Both JAM and DSS results demonstrate good agreement with NPC23 in the high-$z_h$ region but exhibit significant discrepancies at low $z_h$.
Predictions from event generators (JETSET and Pythia8) agree with NPC23 within uncertainties at low and high $z_h$, but deviate in the intermediate region ($0.3 \lesssim z_h \lesssim 0.6$), with Pythia8 showing smaller deviations than JETSET.
When benchmarked against SLD data, a systematic suppression of the experimental results relative to NPC23 predictions for $D_{K^{-}}$ emerges in the intermediate $z_h$ domain ($0.2 \lesssim z_h \lesssim 0.7$). This suppression likely indicates limitations of current FFs, potentially from an overestimated FF from the $s$ quark fragmenting into $K^{-}$ or an underestimated FF from the $u$ quark fragmenting into $K^{+}$ in this kinematic domain. Agreement with SLD measurements is observed within uncertainties at low and high $z_h$.

The observed deviations between state-of-the-art FF predictions and SLD data highlight the potential of precise light-quark jet measurements to provide stringent constraints on fragmentation functions, particularly for strange quarks. Consequently, inclusion of this dataset into global analyses at NNLO accuracy is essential to achieve a more precise and flavor-discriminative determination of fragmentation functions.

\vspace{-0.8em}
\section*{Acknowledgments}
We thank T. Sjostrand for communications on the JETSET predictions, 
Xuan Chen for helpful discussions on the NNLOJET program, and other members of the NPC collaboration for valuable discussions. The work of J.G. is supported by the National Natural Science Foundation of China (NSFC) under Grant No.~12275173, the Shanghai Municipal Education Commission under Grant No.~2024AIZD007, and open fund of Key Laboratory of Atomic and Subatomic Structure and Quantum Control (Ministry of Education).

\section{Data availability}
The data that support the findings of this article are openly available~\cite{BZhou}, embargo periods may apply.

\appendix
\section{The First Term Based on the P2B Method}
\label{Appendix:first}
Through this study, we have developed \texttt{v2.1SLD} of the \texttt{FMNLO} program, which allows for NNLO calculations and grid generation for the hadron energy fraction distribution in light-quark jets and is available from the authors upon request.
Instructions for installation and usage of \texttt{FMNLOv2.1SLD} can be found in Appendix A of Ref.~\cite{Liu:2023fsq}. 
Here, we highlight only the usage of the first term based on the P2B method, which can be calculated using the module \texttt{A4001SLD} within the \texttt{FMNLOv2.1SLD} package.
Importantly, to only consider three light-quark flavors ($u$, $d$, $s$) for the electroweak vertex, one has to modify \texttt{sia/src/inte.f90} located in the main directory \texttt{FMNLOv2.1SLD}. 
That can be achieved by simply replacing $5$ with $3$ at the \texttt{163rd} line of the file.
The parameter card for this module corresponds to the file \texttt{FMNLOv2.1SLD/mgen/A4001SLD/proc.run}, and reads:
\renewcommand{\baselinestretch}{0.78}
\begin{verbatim}
sia A4001SLD
# subgrids with name tags
grid left_1
pdfname 'CT14nlo'
etag 'e-'
htag 'p'
obs 3
zdef 1
cut 0.02
q2d 450.0
q2u 1000.0
xbjd 0.0
xbju 1.0
yid 0.0
yiu 1.0
pdfmember 0
sqrtS 91.18
Rscale 1.0
Fscale 1.0
ncores 30
maxeval 2000000
iseed 11
epol -1.0
ifor 1
nz1 0.15
nz2 0.71
end
\end{verbatim}
The additional parameters compared to similar modules presented in \texttt{FMNLOv2.1}~\cite{Zhou:2024cyk} are as follows: 
\begin{itemize}
\item \texttt{sia} specifies the name of the directory that
 contains the module to be loaded.
\item \texttt{epol} specifies the electron beam polarization. -1.0 denotes a left-handed electron; 1.0 denotes a right-handed electron.
\item \texttt{ifor} specifies the hemisphere to be calculated. 1 selects the quark-tagged hemisphere; -1 selects the antiquark-tagged hemisphere.
\item \texttt{nz1, nz2} specify the lower and upper limits, respectively, for the $z$-component of the signed thrust axis ($n_z$).
\end{itemize}

\section{The Second Term Based on the P2B Method}
\label{Appendix:second}
In \texttt{FMNLOv2.1SLD} we have included a module \texttt{A5005SLD}, dedicated to calculating the second term in Eq.~\eqref{eq:4}.
The parameter card for this module corresponds to the file \texttt{FMNLOv2.1SLD/mgen/A5005SLD/proc.run}, and reads:
\renewcommand{\baselinestretch}{0.78}
\begin{verbatim}
process A5005SLD
# subgrids with name tags
grid left_1
obs 1
cut 0.02
epol -1.0
ifor 1
nz1 0.15
nz2 0.71
# in MG5 format
set ptj 1.0
set lpp1 0
set lpp2 0
set ebeam1 45.59
set ebeam2 45.59
set iseed 11
set muR_over_ref 1.0
set muF_over_ref 1.0
set fixed_ren_scale  True
set muR_ref_fixed 91.18
set req_acc_FO 0.002
end
\end{verbatim}
The additional parameters compared to similar modules presented in \texttt{FMNLOv1.0} are \texttt{epol}, \texttt{ifor}, \texttt{nz1}, and \texttt{nz2}, which are identical to the parameters defined in Appendix~\ref{Appendix:first}.

\section{NLO Calculations as the Benchmark}
\label{Appendix:benchmark}
In \texttt{FMNLOv2.1SLD}, we have included a module \texttt{A1005SLD} for NLO calculation of hadron multiplicity distributions in the energy fraction in light-quark jets at electron-positron collisions.
The parameter card for this module corresponds to the file \texttt{FMNLOv2.1SLD/mgen/A1005SLD/proc.run}, and reads:
\renewcommand{\baselinestretch}{0.78}
\begin{verbatim}
process A1005SLD
# subgrids with name tags
grid left_1
obs 1
cut 0.02
ptj1 0.0
ptj2 50.0
epol -1.0
ifor 1
nz1 0.15
nz2 0.71
# in MG5 format
set ptj 1.0
set lpp1 0
set lpp2 0
set ebeam1 45.59
set ebeam2 45.59
set iseed 11
set muR_over_ref 1.0
set muF_over_ref 1.0
set fixed_ren_scale  True
set muR_ref_fixed 91.18
set req_acc_FO 0.002
# for importance sampling in collinear region
set event_norm bias
end
\end{verbatim}
The additional parameters compared to similar modules presented in \texttt{FMNLOv1.0} are \texttt{epol}, \texttt{ifor}, \texttt{nz1}, and \texttt{nz2}, which are identical to the parameters defined in Appendix~\ref{Appendix:first}.

\bibliographystyle{apsrev4-2}
\bibliography{cite}

\end{document}